\documentclass[aps,pra,twocolumn,showpacs,preprintnumbers,amsmath,
amssymb,superscriptaddress,floatfix,longbibliography,10pt]{revtex4-1}

\usepackage{subfigure}
\usepackage{amsmath}
\usepackage{latexsym}
\usepackage{graphicx}
\usepackage{hyperref}
\usepackage[english]{babel}

\usepackage{graphicx}
\usepackage{dcolumn}
\usepackage{bm}

\newcommand{\var}{\boldsymbol{\rho}}
\newcommand{\vk}{\mathbf{q}}

\begin{document}
\title{Observation of localized multi-spatial-mode quadrature squeezing}
\author{C. S. Embrey}
\affiliation{Midland Ultracold Atom Research Centre, School of Physics and 
Astronomy, University of Birmingham, Edgbaston, Birmingham B15 2TT, UK}
\author{M. T. Turnbull}
\affiliation{Midland Ultracold Atom Research Centre, School of Physics and 
Astronomy, University of Birmingham, Edgbaston, Birmingham B15 2TT, UK}
\affiliation{Gravitational Research Group, Physics Department, John Anderson 
Building, University of Strathclyde, Glasgow, G4 0NG, UK}
\author{P. G. Petrov}
\affiliation{Midland Ultracold Atom Research Centre, School of Physics and 
Astronomy, University of Birmingham, Edgbaston, Birmingham B15 2TT, UK}
\author{V. Boyer}
\affiliation{Midland Ultracold Atom Research Centre, School of Physics and 
Astronomy, University of Birmingham, Edgbaston, Birmingham B15 2TT, UK}

\date{\today}

\begin{abstract}
Quantum states of light can improve imaging whenever the image quality and 
resolution are limited by the quantum noise of the illumination. In the case 
of a bright illumination, quantum enhancement is obtained for a light field 
composed of many squeezed transverse modes. A possible realization of such a 
multi-spatial-mode squeezed state is a field which contains a transverse 
plane in which the local electric field displays reduced quantum fluctuations 
at all locations, on any one quadrature. Using a travelling-wave amplifier, 
we have  generated a multi-spatial-mode squeezed
state and showed that it exhibits localised quadrature squeezing at any point
of its transverse profile, in regions much smaller than its size. We observe
75 independently squeezed regions. The amplification relies on nondegenerate
four-wave mixing in a hot vapor and produces a bichromatic squeezed state.
The result confirms the potential of this
technique for producing illumination suitable for practical quantum imaging.
\end{abstract}

\keywords{Optics, Quantum Physics, Photonics}

\maketitle
\section{Introduction}
When performed with a classical light source, optical measurements are 
limited by the quantum fluctuations of the electromagnetic field, which 
produce noise at the so-called quantum-noise level (QNL). It is however 
possible to improve on the QNL using quantum states of light, for instance
squeezed light~\cite{giovannetti_quantum-enhanced_2004}. To be useful for 
full-field-of-view  
imaging applications, a quantum state of light must be spatially multimode, so 
that it can probe or carry 
spatial information~\cite{kolobov_quantum_2006}. Recently there has been 
substantial progress in few-photon quantum imaging techniques, where the 
illumination is very low and the photons are detected individually. These 
few-photon entangled states have yielded clearer images than those produced by 
the equivalent classical illuminations, whose QNL-limited signal-to-noise 
ratios are nominally poor. In particular, these states have produced images 
of amplitude~\cite{brida_experimental_2010} and 
phase~\cite{ono_entanglement-enhanced_2013,israel_supersensitive_2014} objects 
with noise below the QNL, and interferences displaying better spatial 
resolution~\cite{rozema_scalable_2014}. These experiments do not rely on the 
ability of generating a high level of quadrature squeezing, which remains 
usually very low, but rather on the possibility of generating few photon pairs 
with a good fidelity using low-gain parametric downconversion in a nonlinear 
crystal.

Although very low-level illumination may be required in select applications,
there is a broader interest in applying quantum imaging techniques to the
cases were a bright illumination can be applied. In this case, the 
signal-to-noise ratio at the QNL is much higher and quantum light, specifically
quadrature-squeezed light, can provide an improvement over an already
optimised classical detection. Unlike for the few-photon illumination, quantum
noise reduction with bright illumination is achieved using strong
quadrature squeezing.

The benefit of squeezed light to determine the position of a fixed particle has 
already been demonstrated in a biological environment with a single 
squeezed mode~\cite{taylor_biological_2013}. 
Quantum enhanced imaging of a more complex object can also be achieved by 
squeezing the relevant mode in optical scanning microscopy 
techniques~\cite{taylor_subdiffraction-limited_2014}.
However, improving the spatial resolution in a single-shot imaging of the 
full field of view would require a 
multi-spatial-mode (MSM) quadrature-squeezed light 
field~\cite{kolobov_quantum_2000}. Whilst efficient multimode generation 
of squeezed light has 
been reported in optical parametric oscillators in the time 
domain~\cite{pysher_parallel_2011,roslund_wavelength-multiplexed_2014}, the
realisation of quadrature squeezing in a large number of spatial modes has
remained a long-standing goal in the field of 
quantum optics~\cite{kolobov_spatial_1989}. 
The main hurdle in
the generation of MSM-squeezed light has been the lack of an available strong
multimode nonlinearity. Enhancing weak nonlinearities in $\chi^{(2)}$ crystals
with a spatially degenerate cavity is feasible~\cite{lopez_multimode_2009}, 
and in principle scalable, but
success has been limited to a small number of spatial
modes~\cite{chalopin_direct_2011}. A possible solution is to operate without a
cavity in a pulsed regime, where large peak pump intensities lead to large
levels of squeezing. However this approach has been limited to producing
correlations between twin beams (two-mode squeezed state)~\cite{
jedrkiewicz_detection_2004,brambilla_high-sensitivity_2008,
agafonov_two-color_2010} 
rather than producing a single squeezed beam.
Another workaround is the direct engineering of
overlapping squeezed modes~\cite{treps_quantum_2003} but 
practical scalability is also lacking.

An alternative to parametric downconversion are the large resonant $\chi^{(3)}$
atomic nonlinearities that have been shown to be promising alternatives for the
production of quantum field correlations in the spatial
domain~\cite{boyer_generation_2008,corzo_multi-spatial-mode_2011}.
In this paper, we report on using such nondegenerate four-wave mixing (4WM) in
a hot vapor as a large-gain multimode amplifier. This has allowed us to 
generate a vacuum which is quadrature-squeezed in a large number of spatial 
modes. In particular we have demonstrated localised vacuum quadrature 
squeezing, in a configuration which, when superimposed to a bright coherent 
state, would be suitable for enhanced optical resolution applications.

\section{Background}

In free space, an optical mode is described quantum-mechanically by the field 
quadrature operators $X$ and $Y$. The noncommutativity of $X$ and $Y$ implies 
a Heisenberg inequality $\Delta X \Delta Y \geq \frac{1}{4}$ which is 
responsible for the quantum fluctuations of the electromagnetic field. The 
QNL is reached when the inequality is saturated---it then describes a so-called 
minimum uncertainty state---and the uncertainties on both quadratures 
are equal. It is possible to reduce, or ``squeeze'', the uncertainty on one of 
the quadratures below the QNL, as long as it is compensated by an equal 
increase on the other quadrature. 

\begin{figure}[tb]
    \begin{center}
	\includegraphics[width=0.9\linewidth]{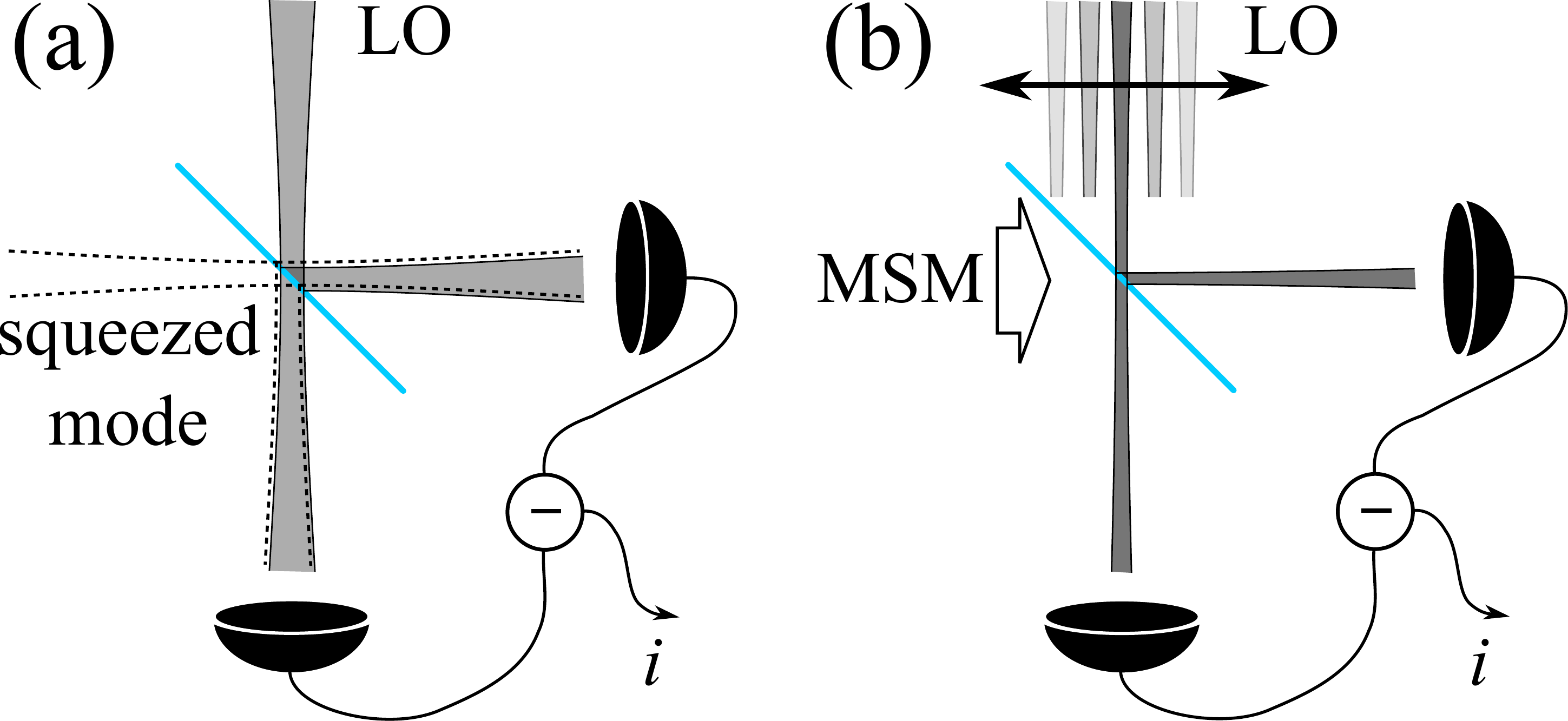}
    \end{center}
    \caption{The homodyne detection of a squeezed state leads to reduced 
noise on the balanced photo-current $i$ below the QNL. (a) For a single 
spatial mode squeezed state the mode of the LO must match the squeezed mode. 
(b) For a hypothetical multi-spatial-mode squeezed state (MSM) the LO could 
have any shape or position.}
    \label{fig:homodyne}
\end{figure}

To illustrate the key signature of a MSM-squeezed state let us first consider
the homodyne detection of a single-mode squeezed state. In such a
configuration, a bright local oscillator (LO) beats with the squeezed mode 
and amplifies the 
fluctuations of one of its quadratures [Fig.~\ref{fig:homodyne}(a)]. Because 
the LO selects the spatial mode to be analysed, it is important to achieve a 
good overlap between the optical modes of the LO and the squeezed field. Soon 
after the first observation of squeezed light~\cite{slusher_observation_1985}
, the question of ``local'' squeezing was raised~\cite{kolobov_spatial_1989}, 
that is to say the possibility of generating and observing a light field with 
reduced quantum fluctuations at any point of its transverse profile. 
Equivalently, such a field would display quadrature squeezing on a homodyne 
detector operated with an arbitrary spatial configuration of the 
LO~\cite{lugiato_improving_1997}, as depicted in Fig.~\ref{fig:homodyne}(b). 
This MSM-quadrature-squeezed field has been theoretically shown to allow an 
improvement of the spatial resolution beyond the QNL in certain schemes of 
optical super-resolution~\cite{kolobov_quantum_2000}, but its efficient
generation has remained elusive until the present work.

To describe more formally the properties of a MSM-squeezed state,
let us consider a light field propagating along the $z$ axis, in a minimum 
uncertainty state, such that in the near field ($z = 0$) the $Y$ quadrature 
is squeezed at all points $\var$ in the transverse plane: $\Delta Y(\var) < 
\frac{1}{2}$. Classically, and in the Fraunhofer diffraction limit, the 
transverse distribution of the far electric field at $z = \infty$ is the 
Fourier transform $E(\vk)$ of the transverse distribution $E(\var)$ of the 
near electric field. Quantum mechanically, this property results in quantum 
correlations in the far field between positions $\vk$ and $-\vk$ due to the 
joint quadratures $X_-(\vk) = [X(\vk) - X(-\vk)]/\sqrt{2}$ and $Y_+(\vk) = [Y(
\vk) + Y(-\vk)]/\sqrt{2}$ being squeezed for all $\vk$ 
(Appendix~\ref{app:correlation}).

Such a state can be created by a travelling-wave amplifier. This device 
creates Stokes and anti-Stokes fields (called twin beams, or probe and 
conjugate, or signal and idler) at the sideband frequencies $-\Omega$ and $
\Omega$ with respect to a central frequency $\omega_0$, which classically 
fulfil the phase conjugation $E(-\Omega) = E^*(\Omega)$
~\cite{boyd_nonlinear_2008}.
  For a thin amplifier diffraction during the propagation is negligible 
and the phase conjugation at the output retains its local character: $E(\var, 
-\Omega) = E^*(\var, \Omega)$ for all $\var$. Quantum mechanically, this 
local phase conjugation translates into local quantum correlations in the 
near field (i.e.\ at the position where they are created in the amplifier). 
Specifically the quantum fluctuations of the joint quadratures $X_-(\var,
\Omega) = [X(\var, \Omega) - X(-\var, \Omega)]/\sqrt{2}$ and $Y_+(\var, \Omega
) = [Y(\var, \Omega) + Y(\var, -\Omega)]/\sqrt{2}$ are reduced below the QNL, 
while the similarly defined joint quadratures $X_+(\var, \Omega)$ and $Y_-(
\var, \Omega)$ are anti-squeezed. When the probe and conjugate fields are 
degenerate, i.e. when $\Omega = 0$, this naturally leads the $Y$ quadrature 
to be squeezed at dc in the near field. 

The amount of squeezing is directly related to the thin amplifier gain
If the gain is too low, a resonant cavity can be used provided it is spatially 
degenerate~\cite{lopez_multimode_2009}. Experimentally this proves 
challenging~\cite{chalopin_direct_2011} and a large gain travelling-wave 
amplifier may be preferable, for instance a 4WM process in a hot atomic 
vapour~\cite{mccormick_strong_2007}.

This nonlinear atomic system has been used to demonstrate symmetric 
correlations in the far field; in a non-degenerate configuration it gives rise 
to entangled images~\cite{boyer_entangled_2008} while in a degenerate 
configuration it produces quadrature squeezing for centrally symmetric 
modes~\cite{corzo_multi-spatial-mode_2011}. The local multimode operation 
of the device was also evidenced by noiseless amplification of near-field 
images~\cite{corzo_noiseless_2012}. Here we report on the direct measurement 
of the local squeezing of a MSM-squeezed state using a homodyne detector with 
an arbitrarily shaped LO, as shown in Fig.~\ref{fig:homodyne}(b).

\begin{figure}[tb]
    \centering
    \includegraphics[width=0.5\linewidth]{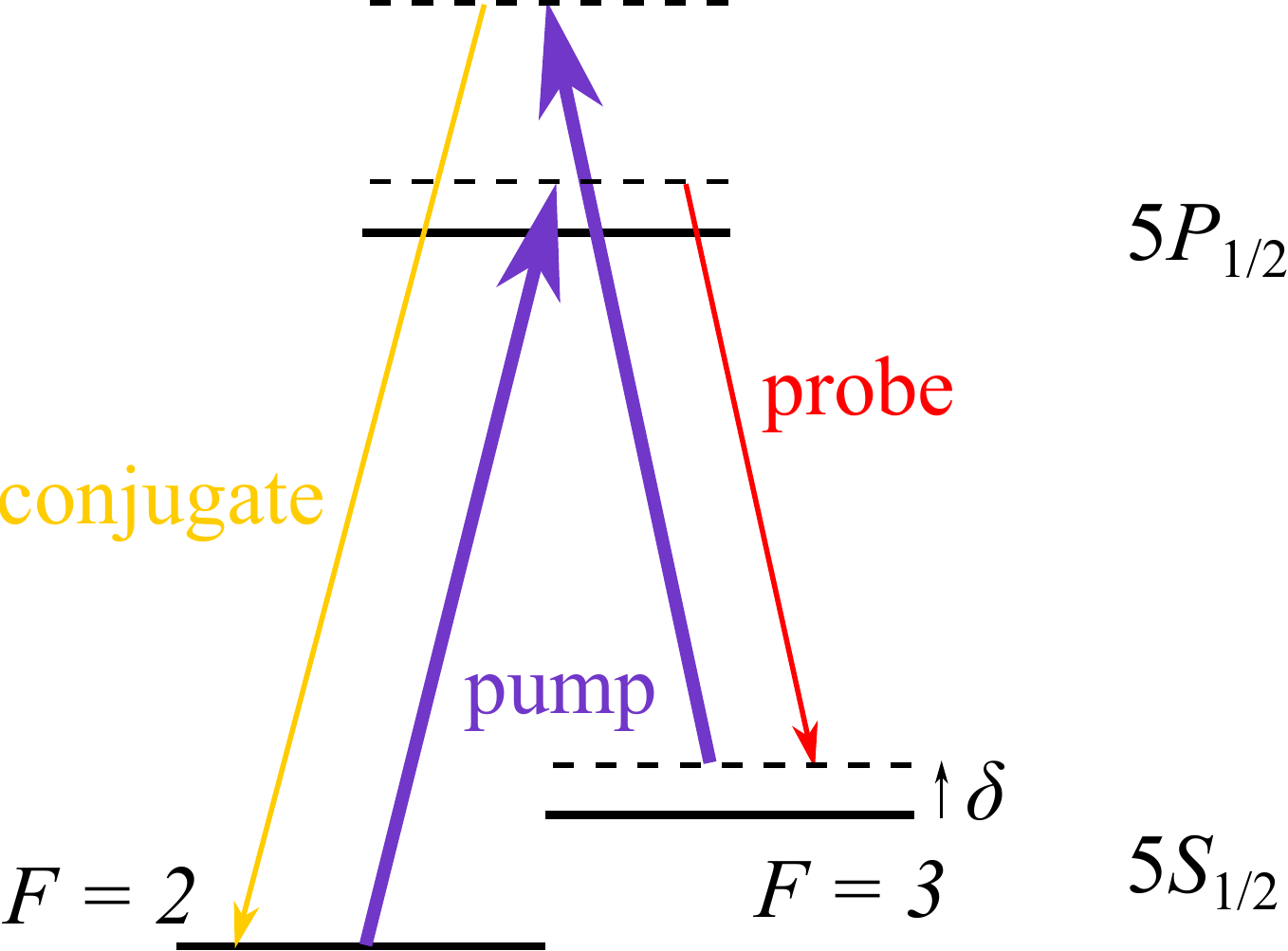}
    \caption{Nondegenerate four-wave-mixing scheme on the $D1$ line of 
$^{85}$Rb. A single pump field creates correlations between the probe 
and conjugate 
fields, whose frequencies are separated by roughly twice the ground-state 
hyperfine splitting.}
    \label{fig:freqs}
\end{figure}

\section{Four Wave Mixing as a Travelling Wave Amplifier}
We use the $D1$ line of rubidium 85 to generate probe and conjugate fields, 
using the 4WM scheme shown in 
Fig.~\ref{fig:freqs}~\cite{mccormick_strong_2007}. A single pump beam, at 
frequency $\omega_0$, couples a probe field, at 
frequency $\omega_p = \omega_0 - \Omega$, with a conjugate field, at 
frequency $\omega_c = \omega_0 + \Omega$, where $\Omega$ is of the order of 
the ground-state hyperfine splitting ($\approx 3$~GHz). The process is 
efficient for a range of detunings $\delta$ between the pump-probe Raman 
transition and the hyperfine splitting. This effectively sets the squeezing 
bandwidth $\Delta\Omega$ to $\approx 20$~MHz.

\begin{figure}[tb]
    \begin{center}
	\includegraphics[width=0.95\linewidth]{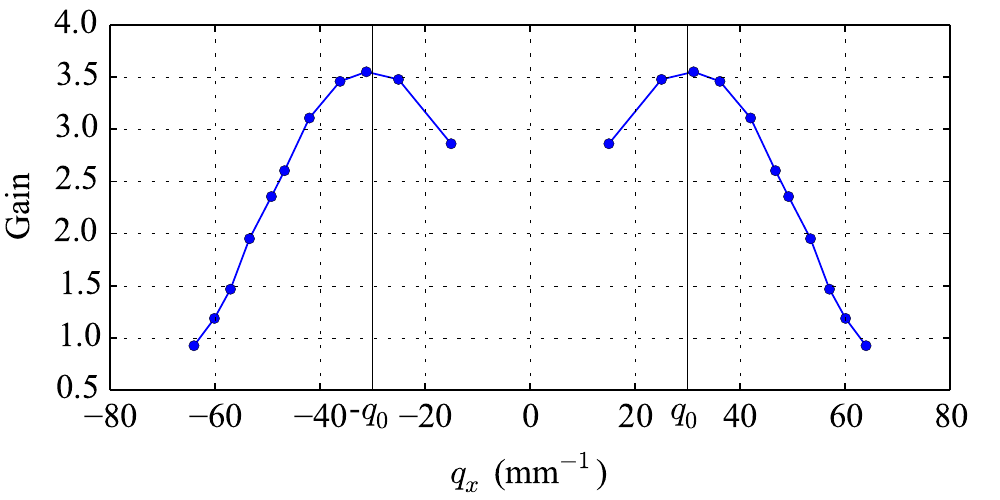}
    \end{center}
    \caption{Spatial gain spectrum as inferred from the spatial gain profile 
in the far field. The probe field is seeded with a Gaussian beam at a 
variable angle with the pump beam and the ratio between the seed power and 
output probe power is measured. The profile has been measured along the $x$ 
direction, but would be the same along any radial direction. The gain at low $
q_x$ is not accurately measurable due to pump light leakage at $q_x=0$.}
    \label{fig:gain-profile}
\end{figure}

\begin{figure}[tb]
    \begin{center}
	\includegraphics[width=0.8\linewidth]{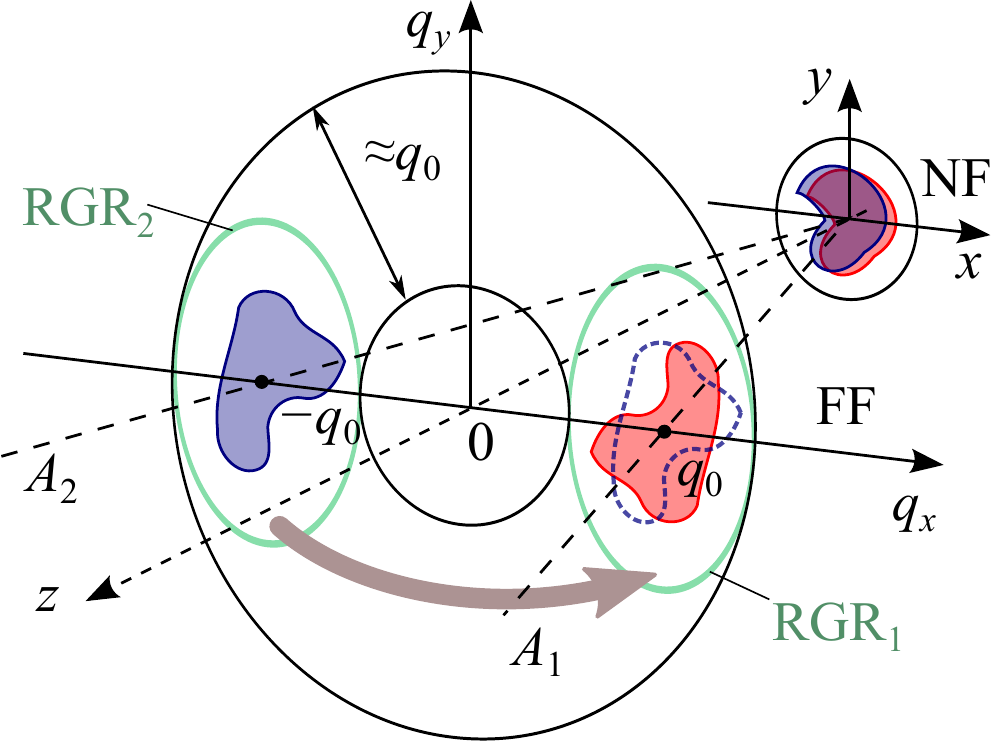}
    \end{center}
    \caption{Geometry of the localised squeezing preparation. A 4WM gain 
medium in the near field (NF) produces local quantum correlations between the 
$\Omega$ and the $-\Omega$ sidebands. As a result, overlapped modes 
fulfilling the phase conjugation relation are correlated (twin mode 1 and 
twin mode 2, represented as overlapped colored regions in the near field). 
After propagation to the far field (FF) those correlated modes are contained 
in a annulus-shaped region resulting from the phase matching condition. On 
this diagram, we only represent the positive-$q_x$ part of twin mode 1 and 
the correlated negative-$q_x$ part of twin mode 2. These correlated modes 
follow the $z$-axial symmetry imposed by the phase-matching condition and we 
further assumed that they are contained in restricted gain regions (RGR) that 
are on the $q_x$ axis. In order to create a multi-spatial-mode squeezed field 
which is fully included in a RGR, i.e.\ in a simply-connected gain region in 
the far field, one can superpose RGR$_2$ on RGR$_1$ on a 50/50 beamsplitter 
(not shown on this diagram) and select $A_1$ as the new optical axis. This is 
equivalent to translating the spectrum of the field in RGR$_2$ by $q_0$ and 
the spectrum of the field in RGR$_1$ by $-q_0$ along $q_x$.}
    \label{fig:cone}
\end{figure}

The phase-matching condition, which requires the probe and conjugate fields 
to propagate symmetrically on opposite sides of the pump, is relaxed by the 
finite length of the rubidium cell. As a result a large number of pairs of 
modes, propagating along slightly different directions, are coupled by the 4WM 
process~\cite{boyer_generation_2008, kumar_degenerate_1994}. Although the 4WM 
follows a co-propagating configuration (forward 4WM), dispersion of the 
index of refraction induces a small angle between the pump axis and the 
direction of maximum probe gain~\cite{turnbull_role_2013}. The resulting far-
field spatial gain profile (Fig.~\ref{fig:gain-profile}) shows that the 
spatial gain spectrum peaks for a finite value of $\left|\vk\right|$ and is 
reduced close to $\left|\vk\right|=0$. Consequently the region of substantial 
gain forms an annulus due to the axial symmetry around the $z$ axis 
(Fig.~\ref{fig:cone}). The gap in the gain around $\left|\vk\right|=0$ means 
that probe 
and conjugate modes with low transverse spatial frequencies are only weakly 
coupled by the 4WM process and cannot develop strong quantum correlations. To 
fix this shortcoming, we can use modes whose probe and conjugate spatial 
frequency spectra are each confined to opposite restricted gain regions 
(RGRs) of the gain annulus. These confined modes see a gapless effective gain 
spectrum in both the $x$ and $y$ directions for both their probe and 
conjugate components (Fig.~\ref{fig:cone}).

In order to avoid the separate propagation of these restricted probe and 
conjugate modes, imposed by the phase-matching condition, we overlap on a 
beam-splitter two correlated propagation axes ($A_1$ and $A_2$) corresponding 
to positions $\pm \vk_0$ in the far field. The direction of $\vk_0$ is 
arbitrarily chosen to be along the $x$ radial direction, as shown in 
Fig.~\ref{fig:cone}. For the matched RGRs to overlap properly the 
magnitude $q_0$ must 
lie close to the middle point of the effective gain spectrum 
(Fig.~\ref{fig:gain-profile}). Redefining the overlapped $A_1$ and $A_2$
 axes as the 
main optical axis the resulting output field is 
\begin{eqnarray}
    E^\prime(\vk, \Omega) &=& \frac{1}{\sqrt{2}}[E(\vk+\vk_0, \Omega) + E(\vk 
- \vk_0, \Omega)], \label{bs-transf1}\label{bs-transf2}
\end{eqnarray}
where the $x$ coordinate of the redefined $\vk$ is restricted to the region [$
-q_0, q_0$]. It can be shown (Appendix~\ref{app:correlation}) that the output 
field exhibits far-field correlations that are symmetric with respect to $\vk_
0$, that is to say with respect to the new optical axis, and that the 
near-field spatial squeezing spectrum is derived from the gapless effective 
spectrum. It therefore contains all the spatial frequencies centred on dc and 
in a bandwidth of the order of $q_0$.

At this stage we have engineered a field with local correlations in the near 
field, which spans a bandwidth $\Delta\Omega=30$~MHz and connects frequency 
sidebands separated by twice the hyperfine splitting, $2\Omega\approx 6$~GHz. 
This composite field forms our squeezed signal.
A homodyne detector using a single-frequency LO at $\omega_0$ would reveal 
the squeezing around an analysing frequency of $\approx3$~GHz. Instead we use 
a bichromatic local oscillator (BLO) as proposed by Marino et 
al.~\cite{marino_bichromatic_2007}, where the single frequency component 
is replaced by two 
frequency components, one for each of the probe and conjugate sidebands.

Since each frequency component is resonant with one of the correlated 
sidebands, the BLO translates the squeezing spectrum from $\approx3$~GHz down 
to dc and the resulting noise on the photo-current $i$ has the similar 
form to that of quadrature squeezing measured by a monochromatic 
LO~\cite{marino_bichromatic_2007}:
\begin{eqnarray}
\left<\Delta i^2\right> & \propto & e^{2s}\cos^2\left(\frac{\chi_p+\chi_c-
\theta_s}{2}\right)\nonumber\\
& &+e^{-2s}\sin^2\left(\frac{\chi_p+\chi_c-\theta_s}{2}\right),
\label{eqn:bi_homo_simple}
\end{eqnarray}
where $\chi_{p,c}$ represents the phase difference between the LO and the 
signal for the probe and conjugate components respectively; $\theta_s$ is the 
squeezing angle; $s$ is the squeezing parameter.

\begin{figure}[tb]
\centering
\includegraphics[width=0.95\linewidth]{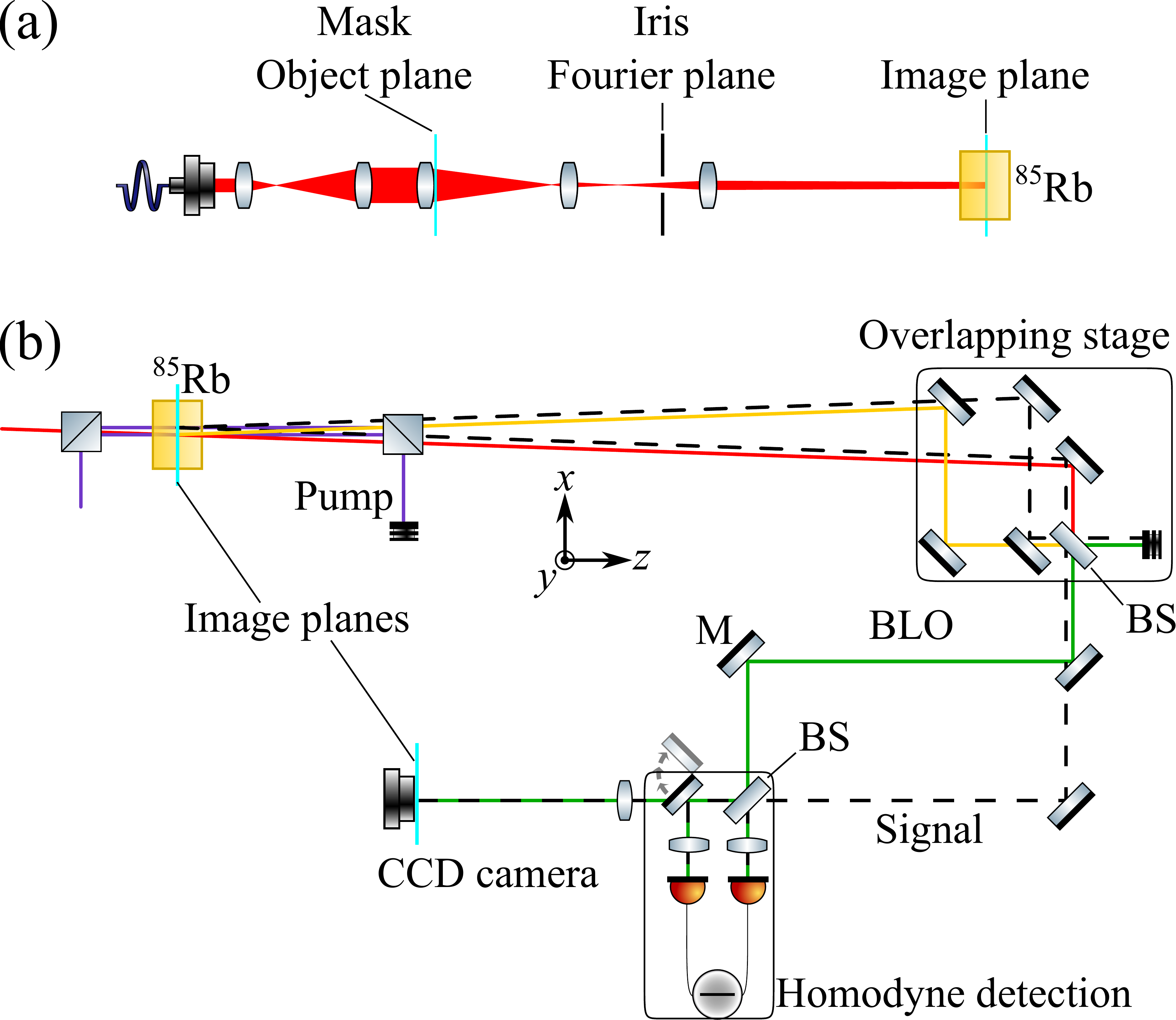}
\caption{Schematic diagram of the setup. (a) shows the path of the seed to 
the vapor cell, with mask and filtering iris positions. (b) shows the 
creation and measurement of the squeezed vacuum. The black dashed lines 
depict the vacuum fields, which at all points contain both the probe and 
conjugate frequencies. The solid lines depict bright fields. The red and 
yellow represent probe and conjugate LO frequencies respectively. The green 
represents the BLO, and the purple the pump field. The cyan lines show the 
mask object and images positions. Where the vacuum and LO fields are slightly 
offset in the diagram they are actually separated vertically in the 
experiment. Nonetheless we use a single beamsplitter for both of them in the 
overlapping stage.}
\label{fig:fullsetup}
\end{figure}

We use a separate 4WM process to generate the required frequency components 
of the BLO. A seed field at the probe frequency in one RGR stimulates the 
generation of bright amplified probe and conjugate fields in opposite RGRs. 
These are superimposed on the overlapping beamsplitter to form the BLO 
(Fig.~\ref{fig:fullsetup}), in a similar manner as for the squeezed signal 
field. This produces a bright bichromatic beam whose two frequency components 
tend 
to propagate along the same axis and have the same mode shape in the near 
field. We will see in section~\ref{sec:phase} that this field has the right 
properties for the BLO.

\section{Experimental Setup}
A simplified experimental setup is shown in Fig.~\ref{fig:fullsetup}. A 
single heated rubidium cell is pumped by a pair of parallel pump beams, thus 
producing two non-overlapping 4WM amplifiers. A set of mirrors and a 
beamsplitter overlap a pair of matched RGRs as in Fig.~\ref{fig:cone}. This 
operation is realised for both 4WM amplifiers. To generate the BLO we seed 
one of the amplifiers, at the probe frequency, with a mode that is contained 
within one of the RGRs. The other amplifier is left unseeded to generate the 
signal field. The resulting BLO and signal fields are fed into a homodyne 
detector to measure the noise on the signal.

\begin{figure}[tb]
\centering
\includegraphics[width=0.95\linewidth]{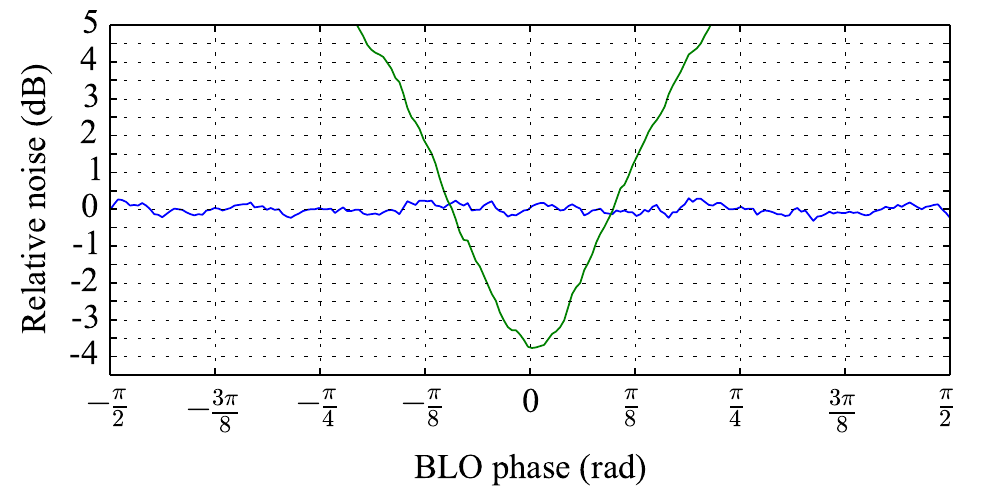}
\caption{Typical squeezing graph, generated by scanning the phase difference 
between BLO and signal fields. For this data the LO pump power is $900$~mW, 
the signal pump power is $950$~mW and the gain is around 4. The electronic 
noise floor can also be subtracted, revealing a squeezing level of 3.8~dB.}
\label{fig:typgraph}
\end{figure}

The relative phase between the signal and LO fields, which controls the 
measured signal quadrature, is tuned by adjusting the optical path length of 
the BLO with a piezo-electric actuator. Using this method we have generated 
squeezing levels of up to $3.6$~dB as show in Fig.~\ref{fig:typgraph}.

Since our main experimental aim is to investigate the local character of the 
quantum correlations, we need to shape the BLO in the near field identically 
for both frequency components. This is achieved by shaping the seed with a 
mask which is optically conjugated with the gain medium 
[Fig.~\ref{fig:fullsetup}(a)]. High spatial frequencies, introduced by the 
mask, are 
filtered out with an aperture located in the Fourier plane. In the same way, 
the position of the BLO in the near field is controlled by steering the seed 
beam before the cell. The actual mode shape and position of the BLO can be 
recorded with an imaging lens located after the overlapping 
beamsplitter. More experimental details can be found in 
Appendix~\ref{app:align}.

\section{Results}
\subsection{Multimode squeezing}

At this stage we have a MSM--quadrature-squeezed field, which should presents 
local quadrature squeezing, and a LO capable of analysing it. The steps 
described above to produce this field are required to remedy issues specific 
to our 4WM process, namely the existence of gaps in the spatial and frequency 
spectra of the gain. Beyond this apparent complexity, the local squeezing is 
the usual consequence of the creation of local correlations inside the 
amplifying medium. The signal field and BLO can be used to realise the simple 
experiment described in Fig.~\ref{fig:homodyne}, that is to say they can 
display squeezing in a homodyne detector arrangement for an arbitrary 
transverse position of the BLO.

We want to show the local character of the squeezing on two perpendicular 
directions, completing the measurement on one direction at a time. To this 
effect we reduce the size of the BLO mode along the direction of interest, 
using a slit as the mask, while allowing the BLO mode to retain its full 
extent in the perpendicular direction.

\begin{figure}[tb]
\centering
\includegraphics[width=0.95\linewidth]{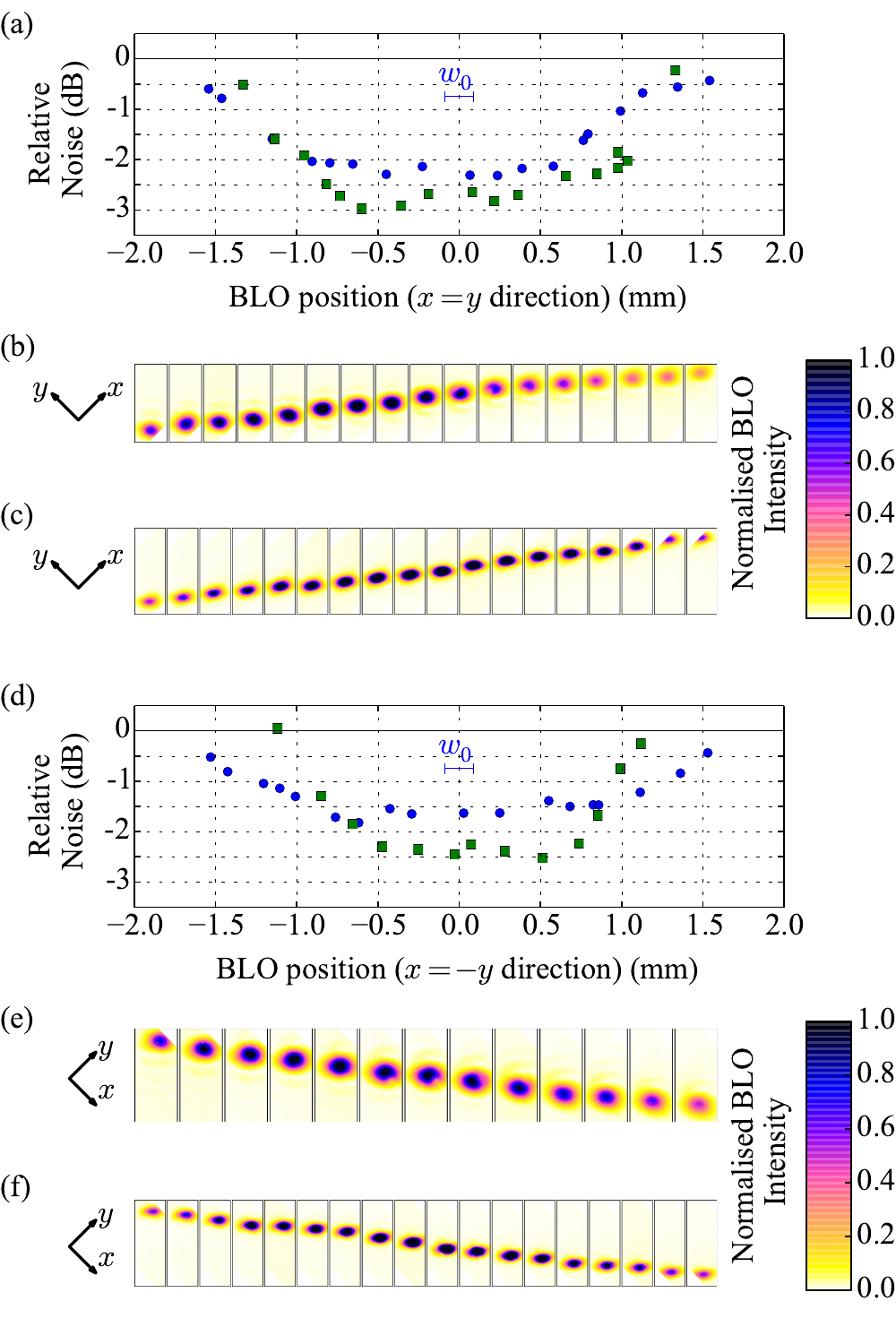}
\caption{Local multimode squeezing. (a) Squeezing as a function of BLO 
position. The BLO position is extracted from (b) and (c) which show images of 
the BLO as it is translated along the $x=y$ direction. The series of images 
(b) (wider BLO) corresponds to the green squares. Series (c) (narrower BLO) 
corresponds to the blue circles. (d) Squeezing as a function of BLO position 
as it is translated along the $x=-y$ direction, again (e) and (f) show the 
images corresponding to the green and blue data respectively. The black lines 
indicate the QNL, the green squares show the data for parameters resulting in 
a gain of 4, with BLO mode waist dimensions of $0.45$~mm by $0.61$~mm, and 
the blue circles show the data for parameters resulting in a gain of 2, with 
BLO mode waist dimensions of $0.31$~mm by $0.58$~mm. All the results are 
corrected for the electronic noise floor (at $-13$~dB). The scale bar 
labelled $w_0$ indicates the size of the coherence area, extracted from 
Fig.~\ref{fig:coharea}}
\label{fig:results}
\end{figure}

The near-field BLO mode shape and the signal quadrature squeezing are 
recorded as the BLO is moved across the near field in the direction of its 
narrow size, whilst keeping its direction of propagation constant. A Gaussian 
fit of the BLO profile gives both its size, which remains constant, and 
position. Figure~\ref{fig:results} shows the degree of squeezing as a 
function of the position of the BLO. The green squares and panels (b) and (e) 
show the squeezing using a gain of around 4. They clearly demonstrate local 
squeezing over a wide range of non-overlapping positions of the BLO in both 
directions and thus the highly-MSM nature of the system.

\begin{figure}[tb]
\centering
\includegraphics[width=0.95\linewidth]{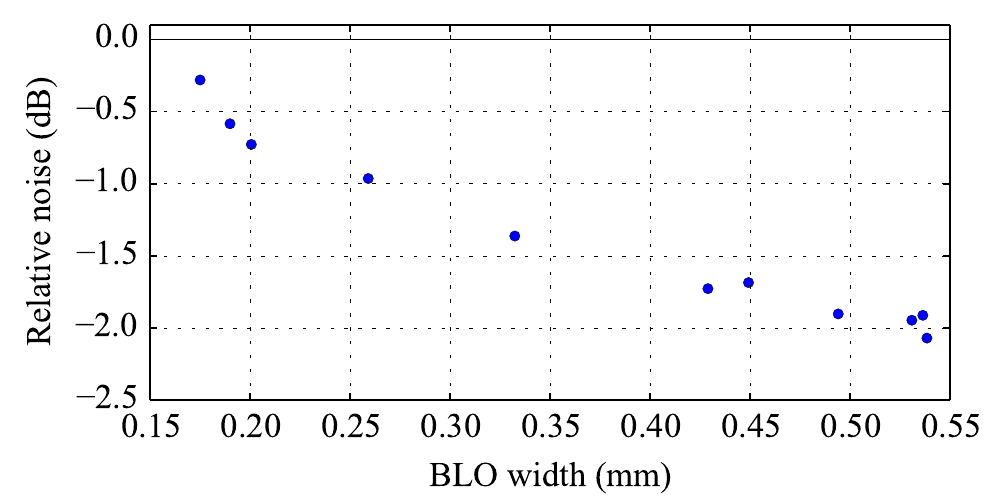}
\caption{Quantum noise reduction as a function of the width of the BLO.}
\label{fig:coharea}
\end{figure}

So far we have assumed a thin medium at $z=0$, in practice the cell has a 
finite length of $12.5$~mm, and propagation effects cannot be fully neglected.  
A mode of very small transverse size will inevitability diffract over the 
length of the gain medium and as a result the correlations cannot be fully 
local. This gives rise to a minimum area over which local squeezing 
can be observed, referred to as the coherence area, and a corresponding 
coherence length~\cite{brambilla_simultaneous_2004}. In the above results we 
have used a BLO with its smaller dimension chosen such that a reasonable 
level of squeezing remains. In the blue circles and panels (c) and (f) in 
Fig.~\ref{fig:results} the gain is reduced to around 2 and squeezing can be 
observed for a smaller slit width, and over a larger range of positions, 
albeit at a lower level. More generally the impact of the slit size on the 
squeezing level can be seen in Fig.~\ref{fig:coharea}. There is a small size $
w_0=0.18$~mm of the BLO for which local squeezing can still be observed. This 
point is reached when the diffracted far-field size of the BLO in the same 
transverse direction occupies the whole RGR.

We extract from Fig.~\ref{fig:results} the size of the squeezing region $l=3.1
$~mm in both the $x=y$ and $x=-y$ directions. Taking $w_0$ to be the 
coherence length one gets a total number of squeezed modes $l^2/4w_0^2=75$.

The measured coherence length can also be compared to the theoretical value 
of the coherence length as described by Lopez et 
al.~\cite{lopez_multimode_2009}. It is the waist of a beam such that the 
Rayleigh range is equal to the 
length of the gain medium, and is given by

\begin{equation}
l_{coh}=\sqrt{\frac{\lambda l_g}{\pi n_s}},\label{eqn:coherence}
\end{equation}
where $n_s$ is the refractive index and is taken to be $1$, $\lambda$ is the 
wavelength and $l_g$ is the length of the gain medium, in this case the 
rubidium cell. With the parameters in our system the theoretical coherence 
length is $l_{coh}=0.056$~mm. The corresponding number of squeezed modes $N$ 
is then calculated by comparing the pump waist $w_p$ and the coherence length:
\begin{equation}
N=\frac{w_p^2}{l_{coh}^2}.\label{eqn:mode_no}
\end{equation}
With our parameters this expression leads to an estimate of 300 independent 
modes being squeezed. However, in our experiment, we only collect and analyse 
a small portion of the 4WM emission annulus (see Fig.~\ref{fig:cone}), and 
thus only have access to a fraction of these modes.

The maximum number of squeezed modes could be increased by enlarging the pump 
beam. Alternatively the same effect could be achieved by reducing the 
coherence length, which is done by reducing the length of the gain medium. 
Both of these adjustments would require the medium to be pumped with a higher 
power in order to attain the same gain.

It can be seen from Fig.~\ref{fig:cone} that the RGRs, as they are formed, 
have different $x$ and $y$ dimensions. We have checked that this results in a 
significant difference in the number of modes between the $x$ and $y$ 
directions.

\subsection{Structure of the LO} \label{sec:phase}
It is clear from the results above that the squeezed field is spatially 
multimode and the BLO can have an arbitrary shape, as long as its spatial 
spectrum fits in the spatial bandwidth of the 4WM process. However in order 
to measure squeezing the probe and conjugate components of the BLO must 
follow the phase conjugation dictated by the 4WM. Classically, for a flat 
pump wavefront, this conjugation reads $E_p(\var) = E_c^*(\var)$ for all $\var
$ in the near field plane. Although this seems a rather straightforward 
condition to fulfil, it should be noted that not matching the BLO to the 
spatial structure of the squeezed vacuum leads to a rapid loss of measured 
squeezing~\cite{la_porta_squeezing_1991}. Indeed squeezing measurements of 
multimode fields are sensitive to LO wavefront distortions. Any imperfection 
in the LO phase profile causes antisqueezed quadratures of higher order 
spatial modes to be measured alongside the squeezed quadrature of the target 
mode, resulting in a noise level which is typically above the QNL.

A possible solution to this experimental difficulty is to use the nonlinear 
process itself to create the LO bright 
field~\cite{kim_quadrature-squeezed_1994,boyer_entangled_2008}. 
We implement this method by stimulating the 
second 4WM process to generate bright probe and conjugate fields that can be 
used to form the BLO (Fig.~\ref{fig:fullsetup}). Provided both pumps have the 
same mode shape, the BLO automatically matches the structure of the squeezed 
vacuum, both in phase and amplitude, irrespective of the chosen pump mode and 
the associated phase conjugation. Note that we still need to accurately 
overlay the very same RGRs for both the BLO and the signal.

In spite of the strong constraints on the wavefront of the LO, it was 
suggested~\cite{lugiato_improving_1997} that spatially multimode squeezing 
can improve the detection of quantum noise reduction due to the relaxed 
constraints on the LO shape. We could indeed verify that the measured 
squeezing was only mildly dependent on the overlap of the BLO with the signal 
in the homodyne detector (e.g.\ tuning of mirror M in 
Fig.~\ref{fig:fullsetup}).

\subsection{Experimental limitations}
There are a number of reasons why a finite amount of squeezing can be 
observed, well below the theoretical value dictated by the gain. The main one 
comes from the way the BLO is generated. Seeding at only the probe frequency 
induces a power imbalance between the probe and conjugate frequency 
components in the BLO, resulting in an uneven detection of the correlated 
sidebands. Increasing the gain reduces this imbalance, but increases the 
antisqueezing in the signal, making the squeezing measurements more sensitive 
to misalignment as explained in the previous section. This trade-off sets the 
optimum gain value in the range of 2--4. The squeezing is also limited by 
other imperfections affecting the reflectance of the mirrors, the 
transmittance of the anti-reflection coatings and the quantum efficiency of 
the detectors.

Throughout this experiment we have chosen to work in the near field where the 
local correlations between probe and conjugate frequency components are 
generated. It is possible to transfer these local correlations to the far-
field. Due to the theoretical axial symmetry of the correlations in the far 
field a flip of one of the RGRs in each of the $q_x$ and $q_y$ directions is 
required (see Fig.~\ref{fig:cone}). In practice the probe and conjugate 
propagate differently due to the Kerr lensing of the probe in the medium, and 
correlated probe and conjugate modes in the far field have slightly different 
shapes~\cite{boyer_entangled_2008}. If one is not concerned with accurate 
control of the LO shape, or sharply localized squeezing, then it is still 
possible to observe multimode squeezing in this fashion. Indeed we have 
successfully measured squeezing up to $2$~dB in this arrangement, with 
results limited by the additional experimental complexity.

\section{Conclusion}

We have demonstrated the generation of a light field which displays local 
squeezing in a total of 75 independent modes using a 4WM system in a hot 
rubidium vapour. The squeezing exists as quantum correlations between distant 
frequency sidebands, however our setup provides a natural way to generate the 
arbitrarily shaped bichromatic local oscillator required to measure the 
multi-spatial-mode squeezing.

Such a quantum state of light can theoretically be used to improve
super-resolution techniques~\cite{kolobov_quantum_2000}, when overlapped with
a bright optical carrier to form a bright illumination. 
In future work, and as a step toward quantum-enhanced
super-resolution, we are aiming to demonstrate local intensity quantum noise
reduction of the resulting illumination in the temporal domain. Directly 
imaging the light with a camera in a series of snapshots should reveal local
intensity fluctuations below the shot noise in arbitrary regions of the
images.

This research was supported by the Engineering and Physical Sciences Research 
Council grants EP/E036473/1 and EP/I001743/1.

\bibliography{multimode}

\appendix
\section{Correlation propagation} \label{app:correlation}
Let us consider the electromagnetic field at the frequency sidebands $\pm 
\Omega$ propagating along the $z$ axis. In a perpendicular plane, taken to be 
the near field at $z=0$, the field operator $E(\var, \Omega)$ can be 
decomposed on the local quadrature operators, themselves expressed as local 
creation and annihilation operators:
\begin{eqnarray}
    X(\var, \Omega) &=& \frac{1}{2}\left[a^\dagger(\var, \Omega) + a(\var, 
\Omega)\right],\\
    Y(\var, \Omega) &=& \frac{i}{2}\left[a^\dagger(\var, \Omega) - a(\var,  
\Omega)\right],
    \label{eq:quadr}
\end{eqnarray}
where $\var$ is the transverse position. The same relations hold in momentum 
space and equivalently, due to Fraunhofer diffraction, in the far field at $z 
= \infty$, the field $E(\vk, \Omega)$ can be decomposed on:
\begin{eqnarray}
    X(\vk, \Omega) &=& \frac{1}{2}\left[a^\dagger(\vk, \Omega) + a(\vk, \Omega
)\right],\\
    Y(\vk, \Omega) &=& \frac{i}{2}\left[a^\dagger(\vk, \Omega) - a(\vk, \Omega
)\right],
    \label{eq:quadk}
\end{eqnarray}
where $a(\vk, \Omega)$ is the spatial Fourier transform of $a(\var, \Omega)$ 
and $a^\dagger(\vk, \Omega)$ is the adjoint of $a(\vk, \Omega)$. Note that $a^
\dagger(\vk, \Omega)$ is also the Fourier transform of $a^\dagger(-\var, 
\Omega)$. In effect, this means that $X(\vk, \Omega)$ is not the Fourier 
transform of $X(\var, \Omega)$.  Physically, this property reflects the phase-
matching condition and as we will now see, it transforms local correlations 
in the near field into symmetric correlations between $\vk$ and $-\vk$ in the 
far field.

A thin nonlinear medium at $z=0$, 
where propagation and the associated diffraction can 
be neglected, creates local correlations which depend on the local phase of 
the pump field.  For instance for a pump with a infinite flat wavefront 
(i.e.\ with a well defined wavevector $\mathbf{k}_0$), the following two joint 
quadratures are squeezed for all $\var$:
\begin{eqnarray}
    X_-(\var, \Omega) &=&  \frac{1}{\sqrt{2}}\left[X(\var, \Omega) - X(\var, -
\Omega)\right], \label{eq:joint1}\\
    Y_+(\var, \Omega) &=&  \frac{1}{\sqrt{2}}\left[Y(\var, \Omega) + Y(\var, -
\Omega)\right].
    \label{eq:joint2}
\end{eqnarray}
In the far field, one can form another joint quadrature and express it as a 
function of $X_-(\var, \Omega)$ and $Y_+(\var, \Omega)$:
\begin{eqnarray}
    X_-(\vk, \Omega) &=& \frac{1}{\sqrt{2}}\left[ X(\vk, \Omega) - X(-\vk, -
\Omega)\right] \label{entang-X}\\
	&=&  \frac{1}{2\sqrt{2}}\left[a^\dagger(\vk, \Omega) + a(\vk, \Omega) \right
. \nonumber\\
	& & - \left. a^\dagger(-\vk, -\Omega) - a(-\vk, -\Omega)\right] \nonumber \\
	&=& \frac{1}{2\sqrt{2}}\, \mathcal{F}\left[a^\dagger(-\var, \Omega) \right. 
\nonumber\\
	& & + \left. a(\var, \Omega) - a^\dagger(\var, -\Omega) - a(-\var, -\Omega)
\right] \nonumber \\
	&=& \frac{1}{\sqrt{2}}\, \mathcal{F}\left[ X_-(\var, \Omega) + X_-(-\var, 
\Omega)\right], \nonumber
\end{eqnarray}
where $\mathcal{F}$ is the Fourier transform operation. Since $X_-(\var, 
\Omega)$ is squeezed for all $\var$ then so is $X_-(\vk, \Omega)$ for all $\vk
$. In a similar fashion, one can show that the same result applies to 
\begin{equation}
    Y_+(\vk, \Omega) = \frac{1}{\sqrt{2}}\left[Y(\vk, \Omega) + Y(-\vk, -
\Omega)\right]. \label{entang-Y}
\end{equation}
This implies that the fields at positions $\pm\vk$ are 
entangled~\cite{duan_inseparability_2000}. 
Ref.~\onlinecite{brambilla_simultaneous_2004} gives an 
account of the near- and far-field correlations for the intensity.

We now show that the overlapping operation shown in Fig.~\ref{fig:cone} 
preserves the local correlations in the near field while restricting the 
accessible spatial spectrum of the fluctuations to the positive side.  By 
overlapping two opposite RGRs of the emission annulus in the far field, one 
translates the fields transversely by $\pm\vk_0$. As a result, the creation 
operator transforms as:
\begin{eqnarray}
    a^\prime(\vk, \Omega) &=& \frac{1}{\sqrt{2}}[a(\vk - \vk_0, \Omega) + a(
\vk + \vk_0, \Omega)], \label{a-sup1}
\end{eqnarray}
with $q_x \in [-q_0, q_0]$. The phase of the superposition is arbitrary and 
has no bearing on the final conclusion. According to Eqs.~(\ref{entang-Y}) 
and (\ref{entang-X}), the entanglement occurs between opposite sidebands $\pm 
\Omega$ and opposite transverse wavevectors $\pm \vk$. For now, we restrict 
ourselves to a particular pair of correlated sidebands, where the $\Omega$ 
sideband comes from the left RGR and the $-\Omega$ sideband comes from the 
right RGR. Within this simplification, the creation operator is:
\begin{eqnarray}
    a^\prime(\vk, \pm\Omega) &=& a(\vk \pm \vk_0, \pm \Omega)
\end{eqnarray}
or equivalently, in the near field:
\begin{eqnarray}
    a^\prime(\var, \pm \Omega) &=& a(\var, \pm \Omega)e^{\mp i \varphi}, 
\label{a-far1}
\end{eqnarray}
with  $\varphi = \vk_0 \cdot \var$. From this we can derive the quadrature 
transformations:
\begin{eqnarray}
    X^\prime(\var, \pm \Omega) &=& X(\var, \pm \Omega)\cos\varphi + Y(\var, 
\pm \Omega)\sin\varphi,~~ \label{quad-far1}\\
    Y^\prime(\var, \pm \Omega) &=& Y(\var, \pm \Omega)\cos\varphi - X(\var, 
\pm \Omega)\sin\varphi,~~ \label{quad-far2}
\end{eqnarray}
and directly we get:
\begin{eqnarray}
    X^\prime_-(\var, \Omega) &=& X_-(\var, \Omega)\cos \varphi - Y_+(\var, 
\Omega)\sin \varphi, \label{joint-near1}\\
    Y^\prime_+(\var, \Omega) &=& X_-(\var, \Omega)\sin \varphi + Y_+(\var, 
\Omega)\cos \varphi. \label{joint-near2}
\end{eqnarray}
Since the joint quadratures $X_-(\var, \Omega)$ and $ Y_+(\var, \Omega)$ are 
locally squeezed, this is also the case for the output  joint quadratures $X^
\prime_-(\var, \Omega)$ and $ Y^\prime_+(\var, \Omega)$. The output field 
also has a contribution from the other possible configuration, where the $
\Omega$ sideband comes from the right RGR and the $-\Omega$ sideband comes 
from the left RGR. This contribution has output joint quadratures that are 
similar to those given in Eqs.~(\ref{joint-near1}) and (\ref{joint-near2}). 
Both contributions are uncorrelated and their noises add in quadrature, so 
that their superposition is also squeezed. As a result the output field 
displays the same local squeezing as the field inside the nonlinear medium, 
while having a continuous spatial spectrum centered on 0 for both the $x$ and 
$y$ directions.

We have considered here a medium of zero length, which results in perfectly
localized squeezing. In practice the cell has a finite length, which
gives rise to a finite minimum size of a squeezed area, called coherence
length. This is studied in section V-A.

\section{Experimental details} \label{app:align}
To ensure relative phase stability, all laser beams are derived from a single 
a Titanium:Sapphire laser, tuned approximately $800$~MHz from the $5^2 S_{1/2}
 \left(F=2\right) \rightarrow 5^2 P_{1/2}$ atomic transition at $795$~nm. The 
main laser beam is split, with two equal parts being used for the LO pump and 
the signal pump at $900$~mW, and a final small portion being used to generate 
the seed beam at the probe frequency. To do this an acousto-optic modulator 
(AOM) is operated at $1.520$~GHz in a double pass arrangement. The seed beam 
has a power of $130~\mu$W. It is amplified by the 4WM, with a gain of around $
4$ with our parameters, and is used to generate a BLO with a total power of 
up to $910~\mu$W. The pump beam has a waist of $1$~mm in the centre of the 
cell, whilst the unvignetted seed beam has a waist of $0.35$~mm. All of the 
noise signals are measured with a spectrum analyser using a detection 
frequency of $1$~MHz, a resolution bandwidth of $100$~kHz and video bandwidth 
of $30$~Hz. To obtain the largest possible squeezing spatial bandwidth the 
parameters are tweaked such that the BLO pump power is $1.3$~W, the signal 
pump power is $580$~mW, and the AOM is operated at $1.523$~GHz. This leads to 
a gain of around 2, and a final BLO power of $215~\mu$W.

A $12.5$~mm-long rubidium vapour cell, heated to $\approx 120^\circ$C, forms 
the gain medium. The cell is contained within a vacuum chamber to avoid the 
convection air currents around the heat pipe, and hence eliminate wavefront 
distortions due to refractive index fluctuations on the signal and BLO 
optical paths.

The production of quadrature squeezing on the overlapping beamsplitter occurs 
only when the phase difference between the two RGRs at the beamsplitter is 
the same for both the probe and the conjugate. We ensure this condition by 
adjusting the difference in the optical path of the two RGRs from the cell to 
the overlapping beamsplitter to an inaccuracy much smaller than the beat 
length between the probe and the conjugate ($5$~cm, corresponding to a 
frequency difference of $\omega_c-\omega_p=6$~GHz). To achieve this we 
temporarily seed the signal 4WM process symmetrically with one seed in each 
RGR, and use the visibility of the resulting bichromatic interference on the 
overlapping beamsplitter to minimise the path length difference. Typically a 
visibility of 99\% can be achieved. Similarly to ensure a good overlap 
between the two frequency components of the LO we use independent 
interferences between each of the components of the LO and the corresponding 
component of the previously aligned seeded signal modes.

To control the size of the BLO in the direction of interest we clip the seed 
with a slit made up of two razor blades. The sharp edges of the slit 
introduce high-order spatial modes, with large $\left|\vk\right|$, lying 
outside of the spatial gain profile. These high-order modes will only be 
present in the probe frequency component. A filtering iris is placed in the 
Fourier plane to remove these spatial frequencies before the 4WM cell. The 
iris size is adjusted to cut at the first zero in the Fourier spectrum.

In order to be able to measure the squeezing using the homodyne detection and 
also image the BLO modes on a camera, a flip mirror is used to control the 
direction of the beam incident on one side of the balanced detector. A single 
lens images the near-field gain region on the camera.
\end{document}